\documentclass[12pt]{article}
\usepackage{url,graphicx,amssymb,amsmath,verbatim,epsfig,epstopdf,natbib,multirow,caption,mathrsfs}
\usepackage{setspace,subfigure,amstext,array,tabu,color}   
\newcolumntype{L}{>{$}l<{$}} 
\usepackage[font=small, margin=1.5cm]{caption}
\usepackage[top=1in, bottom=1.1in, left=1in, right=1in]{geometry}
 
\def\E{{\mathbb E}} 
\def\P{{\mathbb P}} 
\def\Q{{\mathbb Q}} 

\begin{document}
\author{Tim Leung\thanks{Department of Applied Mathematics, University of Washington, Seattle WA 98195. Email:
\mbox{timleung@uw.edu}. Corresponding author.}   
\and Raphael Yan\thanks{Department of Mathematics and Statistics, 
McMaster University, 1280 Main Street West, Hamilton, Ontario L8S 4K1, Canada. Email: \mbox{raphaelyan1218@gmail.com}}  }

\title{A Stochastic Control Approach to \\Managed Futures Portfolios}
\maketitle
\begin{abstract}
We study a stochastic control approach to managed futures portfolios.  Building on the \cite{Schwartz97} stochastic convenience yield model for commodity prices,  we formulate a utility maximization problem for dynamically trading a single-maturity futures or multiple futures contracts over a finite horizon. By analyzing the associated Hamilton-Jacobi-Bellman (HJB) equation, we solve the investor's utility maximization problem explicitly and derive the optimal  dynamic trading strategies in closed form. We provide numerical examples and  illustrate the optimal trading strategies using  WTI crude oil futures data. 
\end{abstract}
 
\newtheorem{theorem}{Theorem}
\newtheorem{proposition}[theorem]{Proposition}
\newtheorem{definition}{Definition}
\newtheorem{remark}{Remark}

\vspace{10pt}
\begin{small}
 
 
\end{small}

\newpage

\section{Introduction}
 Managed futures funds constitute a significant segment in the universe of alternative assets.  These investments are managed by professional investment individuals or management companies known as Commodity Trading Advisors (CTAs), and typically involve trading futures on commodities, currencies, interest rates, and other assets.  Regulated and monitored by both government agencies such as the U.S. Commodity Futures Trading Commission and the National Futures Association,  this class of assets has grown to over  US\$350 billion  in 2017.\footnote{Source: BarclayHedge (\url{https://www.barclayhedge.com}).}  One appeal of  managed-futures strategies   is   their potential to produce  uncorrelated and superior returns, as well as   different risk-return profiles, compared  to the equity market   \citep{CTA1,CTA2}.  While the types of securities traded and strategies are conceivably diverse among managed futures funds, details of the employed strategies   are often unknown.  \cite{demystifying} suggest that momentum-based strategies can help explain the returns of these funds.

 In this paper,  we  analyze  a stochastic dynamic control approach for  portfolio optimization in which the commodity price dynamics and investor's risk preference are incorporated.  The  commodity futures used in our model have the same spot asset but different maturities. Futures with the same spot asset share the same sources of risk. We apply a no-arbitrage approach to construct futures prices from a stochastic spot model. Specifically, we adopt the well-known two-factor model by  \cite{Schwartz97}, which also takes into account the stochastic convenience yield in commodity prices.  We determine the optimal futures trading strategies by solving the associated Hamilton-Jacobi-Bellman (HJB) equations in closed form. The  explicit formulae of our strategies allow for financial interpretations and instant implementation.  Moreover, our  optimal  strategies are explicit functions of the prices of the futures included in the portfolio, but do not require the continuous monitoring of the spot price or stochastic convenience yield. Related to the strategies, we also discuss the corresponding wealth process and certainty equivalent from futures trading. We provide some numerical examples and  illustrate the optimal trading strategies using  WTI crude oil futures data. 
 
 
There is a host of research on the pricing of futures, but relatively few studies apply dynamic stochastic control methods to optimize futures portfolios. Among them, \cite{BichuchShreve}  consider trading a pair of futures but use the arithmetic Brownian motion. In a recent study, \cite{BahmanLeungBasis} study the problem of dynamically trading the price spread between a futures contract and its spot asset under a stochastic basis model. They model  the basis process by a scaled Brownian bridge, and solve a utility maximization problem to derive the optimal trading strategies. These two related studies do not account for  the well-observed no-arbitrage price relationships and  term-structure in the futures market. They motivate  us to consider a stochastic spot model that can generate no-arbitrage futures prices and effectively capture their joint price evolutions.  In our companion paper \cite{firstpaper}, we focus on dynamic pairs trading of VIX futures under a Central Tendency Ornstein-Uhlenbeck no-arbitrage pricing model. All these studies propose a stochastic control approach to futures trading. In contrast, \cite{LeungLiLiZheng2015} introduce an  optimal stopping approach to determine the optimal timing to open or close a futures position under three single-factor mean-reverting spot models. Futures portfolios are also often used to track the spot price movements, and we refer to  \cite{LeungWard,LeungWard_2018} for examples using gold and VIX futures. 
 


The paper is structured as follows.  We describe the futures pricing model and corresponding price dynamics in Section \ref{section2}.  Then in Section \ref{section_u}, we discuss our portfolio optimization problems and provide the solutions in closed form. We also examine the investor's trading wealth process and  certainty equivalence. In  Section \ref{numericalsection}, we provide illustrative numerical results from our model.  Concluding remarks are provided in Section \ref{conclusions}.

\section{Futures Price Dynamics}
\label{section2}
Let us denote the commodity spot price process  by $(S_t)_{t\ge 0}$. Under the \cite{Schwartz97} model, the spot price is driven by a stochastic  instantaneous convenience yield, denoted by $(\delta_t)_{t\ge 0}$ here.
This convenience yield, which was originally used in the context of commodity futures, reflects  the   value  of direct access minus the cost of carry and  can be interpreted as the``dividend yield" for holding the physical asset.  It is the ``flow of services accruing to the holder of the spot commodity but not to the owner of the futures contract" as explained in \cite{Schwartz97}.  

For the spot asset, we  consider its log  price, denoted by $X_t$. Under the \cite{Schwartz97} model, it satisfies the   system of stochastic differential equations (SDEs) under the physical probability measure $\P$:
\begin{eqnarray}X_t &=&\log(S_t),\\
dX_t &=& \left(\mu-\frac{\eta^2}{2}-\delta_t\right) dt + \eta dZ^s_t, \\
d\delta_t &=& \kappa\left(\alpha-\delta_t \right) dt + \bar{\eta} dZ^\delta_t.
\end{eqnarray}
Here, $Z^s_t$ and $Z^\delta_t$ are two standard  Brownian motions under  $\P$ with instantaneous correlation $\rho\in(-1, 1)$. The  stochastic  convenience yield follows the Ornstein-Uhlenbeck model, which is  mean-reverting with a constant equilibrium level $\alpha$, volatility $\bar{\eta}$, and speed of mean-reversion equal to $\kappa$. We require that $ \kappa, \bar{\eta}, \eta >0$  and $\mu, \alpha \in \mathbb{R}$.

The investor's portfolio optimization problem will be formulated  under the  physical measure $\P$, but in order to price the commodity futures we need to work with the risk-neutral pricing measure $\Q$. To this end, we  assume a constant   interest rate $r\ge 0$, and apply  a change of measure from   $\P$ to   $\Q$. The $\Q$-dynamics of the correlated Brownian motions ($Z^s_t, Z^\delta_t$) are given by 
\begin{eqnarray}
d\tilde{Z}^s_t &=&  \frac{\mu-r}{\eta} dt + dZ^s_t \label{s97riskprice1},\\
d\tilde{Z}^\delta_t &=& \frac{\lambda}{\bar{\eta}}dt+dZ^\delta_t \label{s97riskprice2}.
\end{eqnarray}
Consequently, the risk-neutral log spot price evolves according to  
\begin{eqnarray*}
dX_t &=& \left(r-\delta_t-\frac{\eta^2}{2}\right) dt + \eta d\tilde{Z}^s_t, \\
d\delta_t &=& \kappa(\tilde{\alpha}-\delta_t) dt + \bar{\eta} d\tilde{Z}^\delta_t,
\end{eqnarray*}
where we have defined the risk-neutral equilibrium level for the convenience yield by \[\tilde{\alpha}\equiv\alpha-\frac{\lambda}{\kappa}.\] It is adjusted   by the ratio of the market price of risk  $\lambda$ associated with   $Z^\delta_t$  and  the speed of mean reversion  $\kappa$.  With a constant $\lambda$, the convenience yield   again follows the Ornstein-Uhlenbeck model under measure  $\Q$ but with  a different equilibrium level   compared to that under measure  $\P$.

We consider a commodity market that consists of  $n$ traded futures contracts with maturities $T_i, i=1,\ldots,n$.
Let \[F^{(i)}_t\equiv F^{(i)}(t,X_t,\delta_t)= \mathbb{E}[\,e^{X_T}\,|\,X_t, \delta_t\,]\] be the price of the $T_i$-futures at time $t$, which is a function of time $t$,  current log spot price $X_t$, and convenience yield $\delta_t$. For any $i=1,\ldots,n$, the price function $F^{(i)}(t,X,\delta)$ satisfies the PDE
\begin{equation}
\frac{\eta^2}{2}\frac{\partial^2F^{(i)}}{\partial X^2}+\rho \eta \bar{\eta} \frac{\partial ^2F^{(i)}}{\partial X \partial \delta} +\frac{\bar{\eta}^2}{2}\frac{\partial ^2F^{(i)}}{\partial \delta^2}
+\left(r-\delta-\frac{\eta^2}{2}\right)\frac{\partial F^{(i)}}{\partial X}+ \kappa(\tilde{\alpha}-\delta)\frac{\partial F^{(i)}}{\partial \delta}=-\frac{\partial F^{(i)}}{\partial t} \label{fpde},
\end{equation} for 
$(t,x,\delta)\in [0,T_i)\!\times\! (-\infty,\infty)\! \times\! (-\infty,\infty)$, 
where we have compressed the dependence of $F^{(i)}$ on  $(t,X,\delta)$.
The terminal condition is $F^{(i)}(T_i,X,\delta)=\exp(X)$ for $x \in \mathbb{R}$. As is well known  (see \cite{Schwartz97,nfactor}), the futures price  admits the exponential affine form:
\begin{align}
\label{Fit}F^{(i)}_t =  \exp\left(X_t+  A_i(t)+B_i(t)\delta_t\right)\end{align}
for some functions $A_i(t)$ and $B_i(t)$ that depend only on time $t$ and not  the state variables.  The functions $A_i(t)$ and $B_i(t)$ are found from  the   ODEs
\begin{eqnarray} \label{ode1} 
r+\frac{\bar{\eta}}{2}B_i(t)^2+B_i(t)(\alpha\kappa+\rho \eta\bar{\eta})+A_i'(t) = 0, \\
B'_i(t)-\kappa B_i(t)-1 = 0, \label{ode2} 
\end{eqnarray}
for $t\in[0,T_i)$, with terminal conditions $A_i(T_i)=0$ and $B_i(T_i)=0$.
The  ODEs \eqref{ode1} and \eqref{ode2} admit the following explicit solutions: 
\begin{eqnarray}
A_i(t) &=& \left(r-\tilde{\alpha}+\frac{\bar{\eta}^2}{2\kappa^2}-\frac{\eta \bar{\eta} \rho}{\kappa}\right)(T_i-t)  \nonumber \\
&& + \frac{\bar{\eta}^2}{4} \frac{1-e^{-2\kappa(T_i-t)}}{\kappa^3}
+\left(\tilde{\alpha} \kappa + \eta \bar{\eta} \rho -\frac{\bar{\eta}^2}{\kappa} \right) \frac{1-e^{-\kappa(T_i-t)}}{\kappa^2}, \\
B_i(t)&=& -\frac{1-e^{-\kappa (T_i-t)}}{\kappa}. \label{odestar}
\end{eqnarray}

Applying Ito's formula to \eqref{Fit},  the $T_i$-futures price evolves according to  the SDE
\begin{equation} \label{Fsde1} 
\frac{dF^{(i)}_t}{F^{(i)}_t}= \mu_i(t)  dt + \eta  dZ^s_t +  \bar{\eta} B_i(t)  dZ^\delta_t ,\end{equation}
under the physical  measure  $\P$, where the drift is given by 
\begin{eqnarray} 
\mu_i(t) &=& (\lambda+\tilde{\alpha} \kappa +\rho \bar{\eta} \eta) B_i(t)+\frac{\bar{\eta}^2}{2} B_i(t)^2 +\mu+A_i'(t)
+\delta (B'_i(t)-\kappa B_i(t)-1) \\
&=&\mu-r-\frac{\lambda(1-e^{-\kappa(T_i-t)})}{\kappa}.
\label{mu}
\end{eqnarray}
The last equality follows from  \eqref{ode1} and \eqref{odestar}. As a consequence, the  drift of $F^{(i)}_t$ is independent of $X_t$ and $\delta_t$, meaning  that the investor's value function (see \eqref{sup1} or \eqref{sup}) will also be independent of $X_t$ and $\delta_t$. This turns out to be a crucial feature that greatly simplifies  the investor's portfolio optimization problem and ultimately leads to an explicit solution.

To facilitate   presentation,  let us   rewrite the linear combination of $dZ^s_t$ and $dZ^\delta_t$ in \eqref{Fsde1} as
\[\sigma_i(t)  dZ^{(i)}_t \equiv \eta  dZ^s_t +  \bar{\eta} B_i(t)  dZ^\delta_t,\]
where $Z^{(i)}_t$ is a standard Brownian motion and  
\begin{equation}
\sigma_i(t)^2=\eta^2+2\rho \bar{\eta}\eta B_i(t) + \bar{\eta}^2 B_i(t)^2
\label{sigma} \end{equation}
is the  instantaneous volatility coefficient.

Under this model, futures prices are not independent and admit a specific correlation structure. For example, consider the  $T_1$ and $T_2$ contracts. The SDE for the respective futures price is \begin{equation}
\label{Fsde2} 
\frac{dF^{(i)}_t}{F^{(i)}_t}= \mu_i(t)  dt + \sigma_i(t)  dZ^{(i)}_t, \quad i \in \{ 1, 2\},
\end{equation}
The two Brownian motions, $Z^{(1)}_t$ and $Z^{(2)}_t$, are correlated with 
\[dZ^{(1)}_t \, dZ^{(2)}_t =\rho_{12}(t) \, dt.\]
 where 
\begin{equation}
\rho_{12}(t)= \frac{\bar{\eta}^2 B_1(t)B_2(t) +(B_1(t)+B_2(t)) \rho \eta \bar{\eta} + \eta^2}{\sigma_1(t) \sigma_2(t)} 
\label{rho}
\end{equation} is the instantaneous correlation that depends not only on the spot model parameters $(\rho, \eta, \bar{\eta})$ but also the two futures price functions through $B_1(t)$ and $B_2(t)$.

\newpage
\section{Utility Maximization Problem}\label{section_u}
We now present the mathematical formulation for the futures portfolio optimization problem.
To begin, we discuss the case where the investor trades only futures with the same maturity in Section \ref{sect_single}. Then, we extend the analysis to optimize a portfolio with two different futures in Section \ref{sect_pair}.  We will also investigate in Section \ref{sect_ce} the value of trading using the notion of certainty equivalent.  

\subsection{Single-Maturity Futures Portfolio}\label{sect_single}
Suppose that  the investor trades only futures of a single maturity $T_i$ for some chosen $i\in\{1,2,\ldots, n\}$. The trading horizon,  denoted by $T$, must be equal to or shorter than the chosen maturity $T_i$, so we require $T\le T_i$.

We will let $\tilde{\pi}_i(t,F_i)$ denote the number of $T_i$-futures contracts held in the portfolio. The investor can choose the size of the position in the $T_i$-futures, and the position can be long or short at anytime. For brevity, we may write $\tilde{\pi}_i\equiv \tilde{\pi}_i(t,F_i)$.

 

Without loss of generality, we arbitrarily set $i\!=\!1$ in our presentation of the optimization problem and solution. The investor is assumed to  trade only the futures contract and not other risky or risk-free assets.    The   dynamic portfolio consists of $\tilde{\pi}_1(t,F_1)$ units of $T_1$-futures  at time $t$. The self-financing condition means that  the  wealth process satisfies
\begin{equation}
d\tilde{W}_t=\tilde{\pi}_1(t,F^{(1)}_t)  \, dF^{(1)}_t.
\label{wealthprocess1}
\end{equation}
Applying the futures price equations \eqref{Fit} and \eqref{Fsde1}, we can express the  system of SDEs for the wealth process and futures price as 
\begin{equation}
\begin{bmatrix} d\tilde{W}_t \\ d{F}^{(1)}_t  \end{bmatrix}=
\begin{bmatrix} 
{\tilde{\pi}}_1 {\mu}_1(t) {F}^{(1)}_t 
\\ {\mu}_1(t) {F}^{(1)}_t  \end{bmatrix}
dt
+
\begin{bmatrix} 
{\tilde{\pi}}_1 \eta {F}^{(1)}_t & {\tilde{\pi}}_1 \bar{\eta}B_1(t) {F}^{(1)}_t
\\
\eta {F}^{(1)}_t & \bar{\eta}B_1(t) {F}^{(1)}_t\\
\end{bmatrix}
\begin{bmatrix} dZ^s_t \\ dZ^\delta_t \end{bmatrix},
\label{system1}
\end{equation}

\begin{equation}
=
\begin{bmatrix} 
{\tilde{\pi}}_1 {\mu}_1(t) {F}^{(1)}_t 
\\ {\mu}_1(t) {F}^{(1)}_t  \end{bmatrix}
dt
+
\begin{bmatrix} 
{\tilde{\pi}}_1 \sigma_1(t) {F}^{(1)}_t 
\\
\sigma_1(t) {F}^{(1)}_t \\
\end{bmatrix}
dZ^{(1)}_t .
\end{equation}

A control $\tilde{\pi}_1$ is said to be admissible if $\tilde{\pi}_1$ is real-valued progressively measurable, and is such that the system of SDE \eqref{system1} admits a unique solution $(\tilde{W}_t,F^{(1)}_t)$ and  the integrability condition $\E \left( \int_t^T  \tilde{\pi}_1(s,F^{(1)}_s)^2\, (F^{(1)}_s)^2 ds \right) < \infty$ is satisfied. We denote by $\tilde{\mathcal{A}}_t$ the set of admissible strategies  in this case given  an initial investment time    $t$.

The investor's risk preference is described by the exponential utility function \begin{equation}
U(w)=-e^{-\gamma w}, \quad  \text{ for } w\in \mathbb{R},\label{utilexp}
\end{equation}
where $\gamma >0$  is  the constant risk aversion parameter. For a given trading horizon, $[0,T]$, the  investor seeks an admissible strategy  that maximizes the expected utility of terminal wealth at time $T$ by solving the optimization problem
\begin{equation}
\tilde{u}(t,w,F_1)=\sup_{\tilde{\pi}_1\in \tilde{\mathcal{A}}_t} 
\E\left( U(\tilde{W}_T) \,|\, \tilde{W}_t = w, {F}^{(1)}_t = F_1 \right).
\label{sup1}
\end{equation}
We note that the value function is only a function of time $t$, current wealth $w$, and current futures price $F_1$, and does not depend on the current spot price  or convenience yield.

To facilitate presentation, we define the following partial derivatives 
\[
\tilde{u}_t=\frac{\partial \tilde{u}}{\partial t}, 
\quad  \tilde{u}_w=\frac{\partial \tilde{u}}{\partial w},
\quad \tilde{u}_{ww}=\frac{\partial^2 \tilde{u}}{\partial w^2},
\]
\[
\tilde{u}_1=\frac{\partial \tilde{u}}{\partial F_1}, 
\quad \tilde{u}_{11}=\frac{\partial^2 \tilde{u}}{\partial F_1^2},
\quad \tilde{u}_{w1}=\frac{\partial^2 \tilde{u}}{\partial w \partial F_1}.
\]
We expect the value function $\tilde{u}(t,w,F_1)$ to solve the HJB equation
\begin{align}\label{utildehjb1} 
\tilde{u}_t &+ \sup_{\tilde{\pi}_1} \left\{\begin{array}{c} \end{array} \right.
\tilde{\pi}_1 {\mu}_1(t) F_1 \tilde{u}_w  + \tilde{\pi}_1 {\sigma}_1(t)^2 F_1^2 \tilde{u}_{w1} + \frac{1}{2}\tilde{\pi}_1^2 {\sigma}_1(t)^2 F_1^2 \tilde{u}_{ww} \left.  \begin{array}{c} \end{array}\right\} \\\nonumber
&+\frac{{\sigma}_1(t)^2}{2} F_1^2 \tilde{u}_{11} 
+  {\mu}_1(t) F_1 \tilde{u}_1   = 0,
\end{align}
for $(t,w,F_1)\in[0,T)\times \mathbb{R}\times\mathbb{R}_+$, with terminal condition $\tilde{u}(T,w,F_1)= e^{-\gamma w}$ for $(w,F_1)\in \mathbb{R}\times\mathbb{R}_+$. Performing the optimization in \eqref{utildehjb1}, we can express the optimal control $\tilde{\pi}_1^*$  as
\begin{equation}\label{tilpi1}
\tilde{\pi}_1^*(t,F_1)=\frac{\tilde{u}_w \mu_1(t)+ F_1 \tilde{u}_{w1} \sigma_1(t)^2}{F_1 \tilde{u}_{ww}\sigma_1(t)^2}.
\end{equation} 
Substituting this into \eqref{utildehjb1},  we obtain the nonlinear PDE
\begin{equation}
\label{utildehjb1} 
\tilde{u}_t-\frac{\tilde{u}_w^2 \mu_1(t)^2}{2 \tilde{u}_{ww}\sigma_1(t)^2}
-\frac{ F_1 \tilde{u}_w \tilde{u}_{w1} \mu_1(t)}{\tilde{u}_{ww}}
+\frac{F_1(2 \tilde{u}_1 \tilde{u}_{ww} \mu_1(t)-F_1(\tilde{u}_{w1}^2-\tilde{u}_{11} \tilde{u}_{ww})\sigma_1(t)^2)}{2 \tilde{u}_{ww}}=0.
\end{equation}

Next, we conjecture that $\tilde{u}$ depends on $t$ and $w$ only, and apply the transformation
\begin{equation}
\tilde{u}(t,w )=-e^{-\gamma w - \tilde{\Phi}(t )},
\label{utilde}
\end{equation}
  for some function $\tilde{\Phi}(t)$ to be determined. By direct substitution and computation, we obtain the ODE
\begin{equation}
\label{dphitildedt}
\frac{d\tilde{\Phi}}{dt}=-\frac{\mu_1(t)^2}{2\sigma_1(t)^2}
=-\frac{1}{2}\frac{(\lambda(1-e^{-\kappa(T_1-t)}) -\kappa (\mu-r))^2}
{(1-e^{-\kappa(T_1-t)})^2\bar{\eta}^2-2(1-e^{-\kappa(T_1-t)})\kappa \rho \eta \bar{\eta}+\kappa^2\eta^2},
\end{equation}
subject to ${\tilde{\Phi}}(T)$=0. 
In turn, we obtain $\tilde{\Phi}(t)$ by integration
\[
\tilde{\Phi}(t)=\int_t^T \frac{\mu_1(t')^2}{2\sigma_1(t')^2} dt' , \qquad 0\le t\le T.\]

Applying    \eqref{utilde} to \eqref{tilpi1}, we obtain the optimal strategy    
\begin{equation}
\label{pi1star1tilde}
\tilde{\pi}_1^*(t,F_1) = \frac{\mu_1(t)-\sigma_1(t)^2 \tilde{\Phi}_1}{\gamma F_1 \sigma_1(t)^2} = \frac{\mu_1(t)}{\gamma F_1 \sigma_1(t)^2}.
\end{equation}
Using \eqref{odestar}, \eqref{mu}, and \eqref{sigma}, the optimal strategy $\tilde{\pi}_1^*$ in the single-contract case is explicitly given by
\begin{align}
\tilde{\pi}_1^* (t,F_1)=\frac{1}{\gamma F_1}
\frac{\kappa(\lambda(1-e^{-\kappa(T_1-t)}) -\kappa (\mu-r))}
{(1-e^{-\kappa(T_1-t)})^2\bar{\eta}^2-2(1-e^{-\kappa(T_1-t)})\kappa \rho \eta \bar{\eta}+\kappa^2\eta^2}.\label{pi1startildefull}
\end{align}
We observe from  \eqref{pi1startildefull}  that  $\tilde{\pi}_1^*$ is inversely proportional to  $\gamma$ and $F_1$. This means that a  higher risk aversion will reduce the size of the investor's position. A higher futures price will also have the same effect. However, the total cash amount invested in the futures, i.e. $\tilde{\pi}_1^* (t,F_1)F_1$, does not vary with the futures price, and is in fact  a deterministic function of time.  Note that the investor's position  is independent of the equilibrium level of the convenience yield $\alpha$ or $\tilde{\alpha}$, but it depends on the speed of mean reversion $\kappa$, volatility $\bar{\eta}$, and market price of risk $\lambda$ of the convenience yield.

\subsection{Trading Futures of Two Different Maturities}\label{sect_pair}
We now consider the utility maximization problem involving a pair of futures with different maturities. 
Without loss of generality, let $T_1$ and $T_2$  be the two maturities of the futures in the portfolio. The trading horizon $T$ satisfies $T\le \min\{T_1, T_2\}$. The investor continuously trades only  the two futures over time. The trading wealth satisfies the self-financing condition
\begin{equation}
dW_t=\pi_1(t,F^{(1)}_t,F^{(2)}_t)  \, dF^{(1)}_t+\pi_2(t,F^{(1)}_t,F^{(2)}_t)  \, dF^{(2)}_t,
\label{wealthprocess}
\end{equation}
where $\pi_i(t,F^{(1)}_t,F^{(2)}_t)$,   $i = 1, 2$, denote the number of $T_i$-futures held. If it is negative,   the corresponding futures  position is short.
For notational simplicity, we may  write $\pi_i \equiv\pi_i(t,F^{(1)}_t,F^{(2)}_t).$ Writing the trading wealth and two futures prices together in terms of  two fundamental sources of randomness $(Z^{(1)}_t, Z^{(2)}_t)$, we get 
\begin{equation}
\begin{bmatrix} d{W}_t \\ d{F}^{(1)}_t \\ d{F}^{(2)}_t \end{bmatrix}=
\begin{bmatrix} 
{\pi}_1 {\mu}_1(t) {F}^{(1)}_t + 
{\pi}_2 {\mu}_2(t) {F}^{(2)}_t
\\ {\mu}_1(t) {F}^{(1)}_t \\ {\mu}_2(t) {F}^{(2)}_t \end{bmatrix}
dt
+
\begin{bmatrix} 
\pi_1 \sigma_{1}(t) {F}^{(1)}_t& \pi_2 \sigma_{2}(t) {F}^{(2)}_t
\\
\sigma_{1}(t) {F}^{(1)}_t & 0\\
0 & \sigma_{2}(t) {F}^{(2)}_t\\
\end{bmatrix}
\begin{bmatrix} dZ^{(1)}_t \\ dZ^{(2)}_t \end{bmatrix}.
\label{system}
\end{equation}

A pair of controls $(\pi_1,\pi_2)$ is said to be admissible if it is real-valued  progressively measurable, and   such that the system of SDE \eqref{system} admits a unique solution $(W_t,F^{(1)}_t,F^{(2)}_t)$ and the integrability condition $\E \big(\int_t^T  [\pi_i(s,F^{(1)}_s,F^{(2)}_s) F^{(1)}_s]^2 ds \big) < \infty$, for $i=1,2$, is satisfied. We denote by $\mathcal{A}_t$ the set of admissible controls with an initial time of investment $t$.
Next, we define the value function
$u(t,w,F_1,F_2)$ of the investor's portfolio  optimization problem. The investor seeks an admissible strategy  $(\pi_1,\pi_2)$ that maximizes the expected utility from wealth at time $T$, that is,  
\begin{equation}
u(t,w,F_1,F_2)=\sup_{(\pi_1,\pi_2)\in \mathcal{A}_t} \E\left( U(W_T) \,|\, W_t = w, {F}^{(1)}_t = F_1, {F}^{(2)}_t = F_2 \right).
\label{sup}
\end{equation}

\subsubsection{HJB  Equation and Closed-Form Solution}
To facilitate   presentation, we   define the following  partial derivatives  
\[
u_t=\frac{\partial u}{\partial t}, \quad  u_w=\frac{\partial u}{\partial w},\quad 
u_{ww}=\frac{\partial^2 u}{\partial w^2},
\]
\[
u_1=\frac{\partial u}{\partial F_1}, \quad u_{11}=\frac{\partial^2 u}{\partial F_1^2},\quad
u_2=\frac{\partial u}{\partial F_2}, \quad u_{22}=\frac{\partial^2 u}{\partial F_2^2},
\]
 \[
u_{w1}=\frac{\partial^2 u}{\partial w \partial F_1},\quad
u_{w2}=\frac{\partial^2 u}{\partial w \partial F_2},\quad
u_{12}=\frac{\partial^2 u}{\partial F_1 \partial F_2}.
\]
We determine the value function $u(t,w,F_1,F_2)$  by solving the HJB equation
\begin{align}\nonumber 
u_t &+ \sup_{\pi_1,\pi_2} \big[\begin{array}{c} \end{array} 
(\pi_1 \mu_1(t) F_1 + \pi_2 \mu_2(t)  F_2 ) u_w  \\\nonumber 
&+ (\pi_1 \sigma_1(t)^2 F_1^2 +\pi_2{\rho}_{12}(t) \sigma_1(t)\sigma_2(t) F_1 F_2) u_{w1}  + (\pi_2 \sigma_2(t)^2 F_2^2 +\pi_1 \rho_{12}(t) \sigma_1(t)\sigma_2(t) F_1 F_2) u_{w2} \\\nonumber 
&+ \frac{1}{2}(\pi_1^2 \sigma_1(t)^2 F_1^2+ \pi_2^2 \sigma_2(t)^2 F_2^2+\rho_{12}(t) \pi_1\pi_2 \sigma_1(t)\sigma_2(t) F_1 F_2) u_{ww}
   \begin{array}{c} \end{array}\big]+  \mu_1(t) F_1 u_1 +\mu_2(t) F_2 u_2 \\
&+\frac{\sigma_1(t)^2}{2} F_1^2 u_{11} + \frac{\sigma_2(t)^2}{2} F_2^2 u_{22}+\rho_{12}(t) \sigma_1(t) \sigma_2(t) F_1 F_2 u_{12}  = 0, \label{hjb}
\end{align}
for $(t,w,F_1, F_2)\in[0,T)\times \mathbb{R}\times\mathbb{R}_+\times\mathbb{R}_+$, along with the terminal condition
\[{u}(T,w,F_1,F_2)=-e^{-\gamma w}, \quad \text{ for } (w,F_1, F_2)\in  \mathbb{R}\times\mathbb{R}_+\times\mathbb{R}_+.\]

Next, we apply the transformation
\begin{align}\label{utrans}
u(t,w,F_1,F_2)=-e^{-\gamma w -{\Phi}(t,f_1,f_2)},
\end{align}
with  $f_1=\log F_1$ and $f_2=\log F_2$.  Substituting \eqref{utrans} into \eqref{hjb}, we obtain the \emph{linear} PDE for ${\Phi}$:
\begin{align}\nonumber
0 &=\Phi_t+
\left(\frac{1}{2} \frac{{\mu}_1^2}{(1-{\rho}_{12}^2){\sigma}_1^2} 
    + \frac{1}{2} \frac{{\mu}_2^2}{(1-{\rho}_{12}^2){\sigma}_2^2}
-\frac{{\rho}_{12} {\mu}_1 {\mu}_2}{(1-{\rho}_{12}^2){\sigma}_1{\sigma}_2}\right) \\
& ~~+ \frac{{\sigma}_1^2}{2} ({\Phi}_{11}-{\Phi}_1) + \frac{{\sigma}_2^2}{2} ({\Phi}_{22}-{\Phi}_2)
+{\rho}_{12}{\sigma}_1{\sigma}_2 {\Phi}_{12},\label{phipde}
\end{align}
with  $\Phi(T,f_1,f_2)$=0. We have defined the partial derivatives
\[
\Phi_t=\frac{\partial \Phi}{\partial t},\quad
\Phi_1=\frac{\partial \Phi}{\partial f_1},\quad
\Phi_2=\frac{\partial \Phi}{\partial f_2},
\]
 
\[
\Phi_{11}=\frac{\partial^2 \Phi}{\partial f_1^2},\quad
\Phi_{22}=\frac{\partial^2 \Phi}{\partial f_2^2},\quad
\Phi_{12}=\frac{\partial^2 \Phi}{\partial f_1 \partial f_2},
\]
and   suppressed the dependence on $t$, in $\mu_i$, $\sigma_i$, and $\rho_{12}$ to simplify the notation.

We can solve this linear PDE of ${\Phi}$ by using the ansatz
\[{\Phi}(t,f_1,f_2) = a_{11}(t) f_1^2 + a_{1}(t) f_1+a_{22}(t) f_2^2 +a_{2}(t) f_2 +a_{12}(t) f_1 f_2+a(t)\]
to deduce that 
\[a_{11}'(t)=a_{22}'(t)=a_{12}'(t)=0, \text{   } a_{11}(t)=a_{22}(t)=a_{12}(t)=0,\]
\[a_{1}'(t)=a_{2}'(t)=0, \text{   } a_{1}(t)=a_{2}(t)=0.\]
From this, we deduce  that ${\Phi}$ is in fact a function of $t$ only, independent of $f_1$ and $f_2$, and satisfies the first-order differential equation
\[
\frac{d{\Phi}}{dt} = -\frac{{\mu}_1(t)^2 {\sigma}_2(t)^2 +{\mu}_2(t)^2 {\sigma}_1(t)^2-2{\rho}_{12}(t) {\mu}_1(t) {\mu}_2(t) {\sigma}_1(t) {\sigma}_2(t)}{2(1-{\rho}_{12}(t)^2){\sigma}_1(t)^2 {\sigma}_2(t)^2}.
\]
Solving this and applying \eqref{mu}, \eqref{sigma},  and \eqref{rho},  we obtain a closed-form expression for ${\Phi}$. Precisely, 
\begin{equation}\Phi(t)=
\frac{ \left( T-t  \right) 
      \left( {\left( r - \mu  \right) }^2{{\bar{\eta}}}^2 + 
        2\lambda \left( r - \mu  \right) \rho {\bar{\eta}}
         {\eta} + {\lambda }^2\,{\eta}^2 \right) 
      }
{2\left( 1 - {\rho }^2 \right) 
    {{\bar{\eta}}}^2{{\eta}}^2}.
    \label{phiquadratic}
\end{equation}
Applying  \eqref{phiquadratic} to \eqref{utrans}, the value function is given by 
\begin{align}u(t,w) = -e^{-\gamma w - \Phi(t)}.\label{valuefunction2}
\end{align}

Interestingly, as in the single-futures case, the value function is independent of the speed of mean reversion $\kappa$ and equilibrium level $\alpha$ of  the convenience yield process.
Intuitively, it suggests that  the optimal strategy  effectively removes the  stochasticity  of the convenience yield in the investor's maximum expected utility. This feature  is evident again later in the characterization of the optimal wealth process. Moreover, the value function does not depend on the  current futures  prices $(F_1,F_2)$.  The simplicity of the value function is unexpected, especially since there are two stochastic factors and two futures in the trading problem. Nevertheless, it does not mean that the corresponding trading strategies are trivial. In fact, the strategies depend not only on other model parameters but also the futures prices, as we will discuss next.

By applying  \eqref{utrans} and \eqref{phiquadratic} to  \eqref{hjb},  we obtain the optimal trading strategies
\begin{align}
\pi_1^* (t,F_1,F_2) 
&=  \frac{1}{\gamma (1-{\rho}_{12}(t)^2){\sigma}_1(t) F_1} \left( \frac{{\mu}_1(t)}{{\sigma}_1(t)}-{\rho}_{12}(t)\frac{{\mu}_2(t)}{{\sigma}_2(t)} \right), \label{pi1star} \\
\pi_2^* (t,F_1,F_2) 
&=  \frac{1}{\gamma (1-{\rho}_{12}(t)^2){\sigma}_2(t) F_2} \left( \frac{{\mu}_2(t)}{{\sigma}_2(t)}-{\rho}_{12}(t)\frac{{\mu}_1(t)}{{\sigma}_1(t)} \right). \label{pi2star}
\end{align}
In this case with two futures,  for either $i=1,2$, the corresponding optimal strategy  $\pi_i^*$  is a function of $F_i$, but does not depend on the price of the other futures $F_j$, for $i \! \ne \! j$. Also note that if $\rho_{12}(t)$ is zero, 
then the two-futures strategy reduces to the single-futures strategy, as in 
\eqref{pi1star1tilde}, which is given explicitly by \eqref{pi1startildefull}.

We recall  \eqref{mu}, \eqref{sigma}, and \eqref{rho}, and express the optimal strategies explicitly in terms of model parameters. Precisely,

\begin{equation}
\pi_1^* =-
\frac{e^{\kappa \left( T_1-t \right) }
    \left( \left( e^{t\kappa } - e^{\kappa {{T }_2}} \right)
         \left( r - \mu  \right) {{\bar{\eta}}}^2 + 
      \left( e^{t\kappa }\lambda  + 
         e^{\kappa {{T }_2}}
          \left( r\kappa  - \lambda  - \kappa \mu  \right) 
         \right) \rho {\bar{\eta}}{\eta} + 
      e^{\kappa {{T }_2}}\kappa \lambda 
       {{\eta}}^2 \right) }{F_1 \left( e^
       {\kappa {{T }_1}} - e^{\kappa {{T }_2}} \right) 
     \gamma \left( 1 - {\rho }^2 \right) {{\bar{\eta}}}^2   \eta^2},\label{pi1}
\end{equation}
\begin{equation}\pi_2^*=
\frac{e^{\kappa \left( T_2-t  \right) }
      \left( \left( e^{t\kappa } - e^{\kappa {{T }_1}}
           \right) \left( r - \mu  \right) {{\bar{\eta}}}^2 + 
        \left( e^{t\kappa }\lambda  + 
           e^{\kappa {{T }_1}}
            \left( r\kappa  - \lambda  - \kappa \mu  \right) 
           \right) \rho {\bar{\eta}}{\eta} + 
        e^{\kappa {{T }_1}}\kappa \lambda 
         {{\eta}}^2 \right) }{F_2\left( e^
         {\kappa {{T }_1}} - e^{\kappa {{T }_2}}
        \right) \gamma \left( 1 - {\rho }^2 \right) {{\bar{\eta}}}^2 \eta^2}. \label{pi2}
\end{equation}
Thus we see that the optimal controls $\pi_1^*$ and $\pi_2^*$ do not depend on the current spot price $S_t$ or convenience yield $\delta_t$, and is also independent on the equilibrium of the convenience yield  $\alpha$. For practical applications, this independence removes the burden to estimate or continuously monitor the spot price or convenience yield. Nevertheless,  the optimal controls do depend on all the other parameters, namely $\mu,r,\kappa,\eta,\bar{\eta},\rho,\mbox{ and } \lambda$. Lastly, we notice from  \eqref{pi1star} that, when ${\rho}_{12}(t)$ (see \eqref{rho}) equals zero, $\pi_1^*$ in this    two-futures case    is identical to     $\tilde{\pi}_1^*$ from the single-futures case (see \eqref{pi1star1tilde}).

\begin{remark}
Naturally, one can consider trading futures with  more than two  maturities. However, in such case under the Schwartz two-factor model,  there is an infinite number of solutions to the corresponding utility maximization problem and the additional futures are redundant, as we show in  Appendix \ref{append}.
\end{remark}

\subsubsection{Optimal Wealth Process}
To derive the optimal wealth process, we substitute the optimal futures positions, ${\pi}_1^*$ and ${\pi}_2^*$, into the wealth equation \eqref{wealthprocess} and get 
\begin{eqnarray*}
dW_t&=&\pi_1^* dF^{(1)}_t+\pi_2^* dF^{(2)}_t \\
&=& \mu_W dt 
+ (\pi_1^* F^{(1)}_t +\pi_2^* F^{(2)}_t)\eta  dZ^s_t 
+ (\pi_1^* F^{(1)}_t B_1(t) +\pi_2^* F^{(2)}_t  B_2(t) ) \bar{\eta} dZ^\delta_t\\
&\equiv& \mu_W dt +\sigma_W dZ^W_t,
\end{eqnarray*}
where   we have defined
\begin{align}
\mu_W &= \pi_1^* F^{(1)}_t \mu_1(t)+\pi_2^* F^{(2)}_t \mu_2(t)\notag\\
&= \frac{  {\left( r - \mu  \right) }^2{{\bar{\eta}}}^2 + 
        2\lambda \left( r - \mu  \right) \rho {\bar{\eta}}
         {\eta} + {\lambda }^2\,{{\eta}}^2  }
{\gamma\left( 1 - {\rho }^2 \right) {{\bar{\eta}}}^2{{\eta}}^2}\label{muw}
\end{align}
and
\begin{eqnarray}
\sigma_W^2 &=& (\pi_1^* F^{(1)}_t +\pi_2^* F^{(2)}_t)^2 \eta^2 + (\pi_1^* F^{(1)}_t B_1(t) +\pi_2^* F^{(2)}_t  B_2(t) )^2 \bar{\eta}^2  \notag\\
&&+2\rho \eta  \bar{\eta} (\pi_1^* F^{(1)}_t +\pi_2^* F^{(2)}_t) (\pi_1^* F^{(1)}_t B_1(t) +\pi_2^* F^{(2)}_t  B_2(t) ) \notag \\
&=& \frac{  {\left( r - \mu  \right) }^2{{\bar{\eta}}}^2 + 
        2\lambda \left( r - \mu  \right) \rho {\bar{\eta}}
         {\eta} + {\lambda }^2\,{{\eta}}^2  }
{\gamma^2 \left( 1 - {\rho }^2 \right) {{\bar{\eta}}}^2{{\eta}}^2} \notag\\
&=& \frac{\mu_W}{\gamma}.\label{sigmaw}
\end{eqnarray}
In \eqref{muw} and \eqref{sigmaw} we have  used  \eqref{pi1} and \eqref{pi2}.

Note that both $\mu_W$ and $\sigma_W$ are constant. This implies that the wealth process, under the optimal trading strategy, is an arithmetic Brownian motion with constant drift and volatility. Moreover, these two constants do not depend on the speed of mean reversion $\kappa$ and equilibrium level $\alpha$ of the convenience yield process.  This is why the value function is also independent of these two parameters. The financial intuition is that the optimal   strategy suggests trading in a way that removes the randomness stemmed from  the convenience yield process. As a special case, when  $\mu=r$ and $\lambda=0$,  the $\P$ measure is identical to $\Q$. This will lead to $\pi^*_i=0, i=1,2$, and in turn a constant wealth, with $\mu_W =\sigma_W=0$.


\subsection{Certainty Equivalent}
\label{sect_ce}
Next, we consider the \emph{certainty equivalent} associated with the trading opportunity in the futures. The certainty equivalent is the cash amount that derives the same utility as the value function. First, we consider the single-futures case.  Recall from \eqref{utilexp} and \eqref{utilde}  that the investor's utility and value functions are both of exponential form. Therefore, the certainty equivalent is given by
\begin{equation}
\label{cetilde}
\tilde{C}^{(i)}(t,w)\equiv U^{-1}( \tilde{u}(t,w)) =   w+\frac{\tilde{\Phi}^{(i)}(t)}{\gamma}.
\end{equation}
 Here, the superscript $(i)$ refers to the  futures with maturity $T_i$  in the portfolio.
From \eqref{cetilde}, we observe that the certainty equivalent is the sum of the investor's wealth $w$ and the time-deterministic component  ${\tilde{\Phi}^{(i)}(t)}/{\gamma}$, which is  positive and inversely proportional to  the risk aversion parameter $\gamma$. All else being equal, a more risk averse investor has a lower certainty equivalent, valuing the futures trading opportunity less.  Interestingly, the  certainty equivalent  does not depend on the   current futures  prices $F_1$ but it does depend on the model parameters that appear in  the futures price dynamics. 

Similarly, the certainty equivalent from dynamically trading two futures with different maturities is given by 
\begin{equation}
C(t,w) \equiv U^{-1}(u(t,w))=w+\frac{\Phi(t)}{\gamma},
\label{ce}
\end{equation}
where $u(t,w)$ is the value function in \eqref{valuefunction2} and  $\Phi$ is given by \eqref{phiquadratic}.

Since the certainty equivalents in both the single-futures and two-futures cases have the same linear dependence on wealth $w$,  we will for simplicity set   $w=0$ in our numerical examples to compare across these cases.  To this end, we denote $\tilde{C}^{(i)}_0(t) \equiv\tilde{C}^{(i)}(t,0)$ and $C_0(t) \equiv C(t,0)$.  

%

\section{Numerical Implementation}
\label{numericalsection}

We   now    examine   our model through a number of   numerical examples using  simulated and empirical data.  For our examples, we will use the estimated parameters values found in  \cite{Ewald2018}. They are displayed here in Table~\ref{s97table}.
The drift parameter $\mu$ of the spot price was not given in \cite{Ewald2018}, so we set $\mu=1\%$ for our examples. 
We use federal funds rate as a proxy for the instantaneous interest rate $r$ which,
during the calibration period, hovered around $0.1\%$.\footnote{Data from www.macrotrends.net.} The default value for the risk aversion coefficient $\gamma$ is $1\%$ unless noted otherwise.
 
\vspace{5pt}
\begin{table}[h]
\centering
{\setlength\extrarowheight{5pt}
$\begin{tabu}{|c|c|c|c|c|c|c|}
\hline
\mu & \kappa & \eta & \bar{\eta} & \rho & \lambda & r \\ \hline 
0.010 & 0.800 & 0.450 & 0.500 & 0.750 & 0.050 & 0.001 \\ \hline
\end{tabu}$}\caption{The \cite{Schwartz97} model parameters estimated by \cite{Ewald2018}.}
\label{s97table}
\end{table}
\vspace{5pt}

\begin{figure}[h]
    \centering
{\includegraphics[width=4.2in]{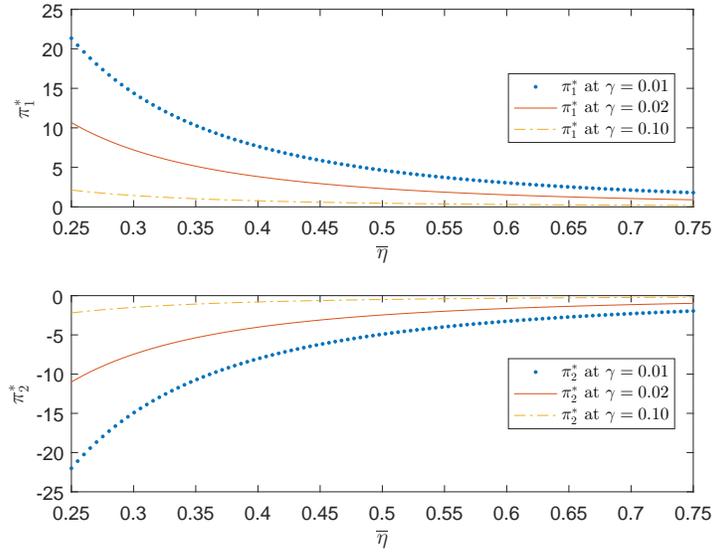}}
\caption{
Optimal positions,  $\pi_1^*$ and ${\pi}_2^*$, respectively in the $T_1$-futures and $T_2$-futures in the two-futures case plotted for $\bar{\eta}\in[0.25, 0.75]$, at three levels of risk aversion $\gamma$. Common parameters are displayed in Table~\ref{s97table}, with  $F_1=100$ and $F_2=100$. }
\label{PLOT1}
\end{figure}

In Figure~\ref{PLOT1}, we show the dependence of the optimal positions,   $\pi_1^*$ and ${\pi}_2^*$, respectively in the $T_1$-futures and $T_2$-futures in the two-futures case on  the  volatility parameter  $\bar{\eta}$ of the convenience yield process, for three different  risk aversion levels. 
Observe that ${\pi}_1^*$ at all three levels of $\gamma$ is positive and decreasing in $\bar{\eta}$ while  $\pi_2^*$ is negative and increasing in $\bar{\eta}$.  
With the parameters given in Table~\ref{s97table}, we are long the $T_1$-futures $F^{(1)}$ and short the $T_2$-futures $F^{(2)}$.
When we rearrange the formulae \eqref{pi1} and \eqref{pi2}  for $\pi_1^*$ and ${\pi}_2^*$, respectively, and collect terms involving $\bar{\eta}$,
we see that for both $i=1,2$, the optimal strategies  are of the form $A_i+B_i/\bar{\eta}+C_i/\bar{\eta}^2$, 
which means that the absolute value   of the each strategy  $\pi_i^*$   decreases as $\bar{\eta}$ increases, with other variables held constant.
The practical  consequence is that the number of contracts held, on both the long and   short sides, are decreasing as the volatility of the stochastic convenience yield process $\delta_t$  increases.
This is in line with a risk-averse trader's intuition that  less exposures on both legs of the traded pair should be preferred, if the volatility of the stochastic  convenience yield  is high.
Furthermore, the positions increase in size 
(more positive for $\pi_1^*$ and more negative for $\pi_2^*$) 
as risk aversion decreases. This is obvious given the inverse relationship 
between $\gamma$ and $\pi_i^*$ as seen in Eq~\eqref{pi1star} and \eqref{pi2star}.

\begin{figure}[h]
    \centering
{\includegraphics[width=4.2in]{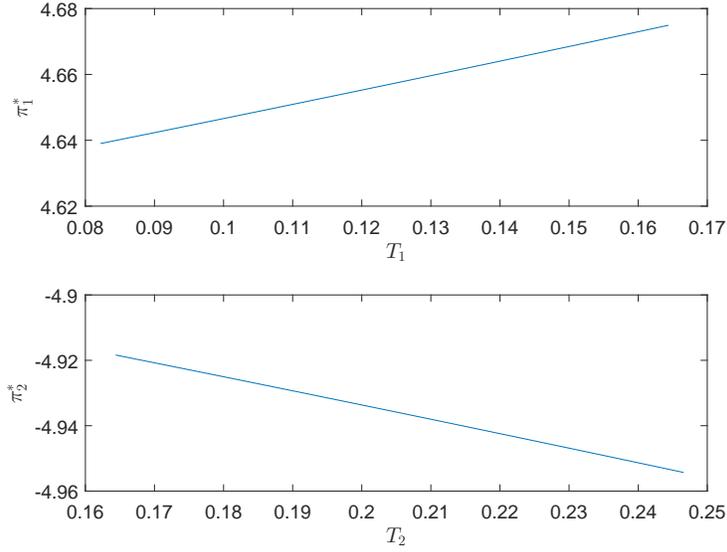}}
\caption{Optimal positions,  $\pi_1^*$  in the $T_1$-futures and    ${\pi}_2^*$ in the  $T_2$-futures in  the two-futures portfolio, plotted as  a function of $T_1$ and $T_2$ respectively, with parameters as displayed in Table~\ref{s97table}, and $F_1=100$ and $F_2=100$.
}
\label{PLOT2}
\end{figure}

Figure~\ref{PLOT2} illustrates how the optimal futures positions, $\pi_1^*$ and $\pi_2^*$, vary with respect to maturity. First of all, the two positions are of different signs and their sizes are very close.   As  maturity $T_1$ or $T_2$ lengthens, the size of the corresponding futures  position   increases, with $\pi_1^*$ becoming more positive and $\pi_2^*$ more negative. However, the change is very small as the scale on the y-axis shows, so one can interpret this as the positions are not very sensitive to the futures maturities.

\begin{figure}[h]
    \centering
{\includegraphics[width=4.2in]{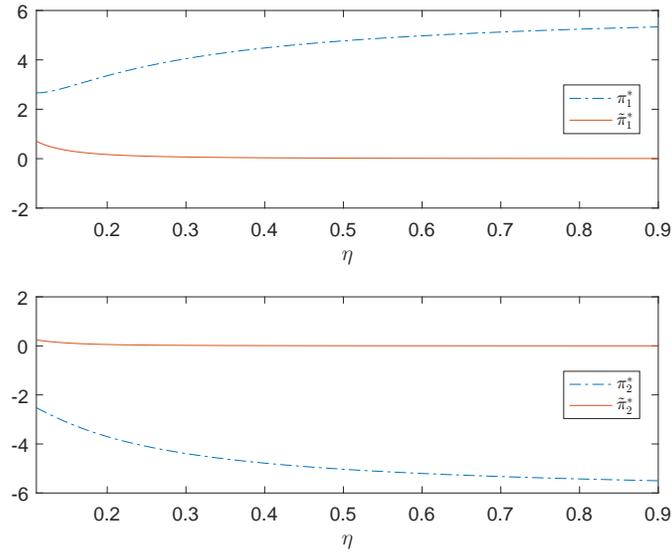}}
\caption{Optimal futures position ${\pi}_i^*$ (dashed) in the 2-contract portfolio  and $\tilde{\pi}_i^*$ (solid)  in the single-contract portfolio (with the $T_i$-futures) plotted over $\eta \in[0.1,0.9]$. Parameters are taken from Table~\ref{s97table}, with  $F_1=100$ and $F_2=100$.
}
\label{PLOT3}
\end{figure}
\clearpage

In Figure \ref{PLOT3} we compare the optimal trading strategies, $\pi_1^*$ and $\pi_2^*$ for two futures to   the optimal strategy $\tilde{\pi}_i^*$ for trading a single futures. We  plot the strategies as functions of $\eta$, the volatility of the spot price, using same set of parameters as in Table~\ref{s97table}.
When trading a single contract, the corresponding optimal strategy, 
$\tilde{\pi}_1^*$ and $\tilde{\pi}_2^*$, 
are both very small near zero. 
However, it can be seen that they do increase slightly in size when $\eta$ becomes small, as volatility decreases.
 
This is in contrast to the two-contract case where the optimal strategies are 
$\pi_1^*$ and $\pi_2^*$. Both increase, in opposite directions, as $\eta$ increases.
This shows that despite the increase in risk as $\eta$ increases,
paired positions in $\pi_1^*$ and $\pi_2^*$, of opposite signs, 
will increase as volatility of the spot process increases.

It is also interesting to note the size of the positions in the single contract cases as compared to the pair-trading case. 
When we are constrained to trade only single contracts, 
that is when the admissible set is $\tilde{\mathcal{A}}_t$ as opposed to $\mathcal{A}_t$, the position is much smaller. 
Under the current model, the presence of multiple contracts of different maturities significantly increases trade volume and allows the trader to take much bigger hedged trades.

\begin{figure}[h]
    \centering
{\includegraphics[width=4.2in]{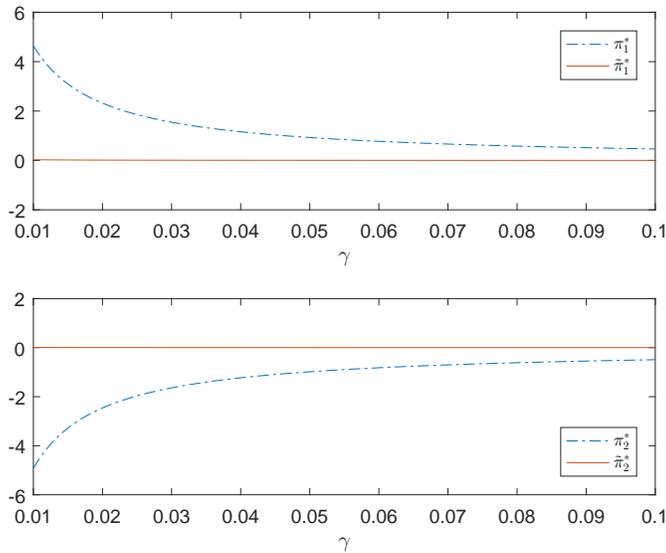}}
\caption{Optimal futures position ${\pi}_i^*$ (dashed) in the 2-contract portfolio  and $\tilde{\pi}_i^*$ (solid)  in the single-contract portfolio (with the $T_i$-futures) plotted over $\gamma \in[0.01,0.1]$. Parameters are taken from Table~\ref{s97table}, with  $F_1=100$ and $F_2=100$.
}
\label{PLOT4}
\end{figure}

In Figure~\ref{PLOT4} we plot the optimal strategies as functions of $\gamma$,
the risk aversion coefficient. 
Obviously, given the inverse relationship between 
$\gamma$ and $\pi_i^*$ as seen in Eq~\eqref{pi1star} and \eqref{pi2star},
as well as between
$\gamma$ and $\tilde{\pi}_i^*$ as seen in Eq~\eqref{pi1star1tilde},
the optimal positions are expected to decrease in magnitude.
What is interesting to note is the insensitivity of 
$\tilde{\pi}_i^*$ with respect to $\gamma$,
in comparison to $\pi_i^*$.
This means that in the single futures case, 
the position will be small regardless of the level of risk aversion.

\begin{figure}[h]
    \centering
\includegraphics[width=4.2in]{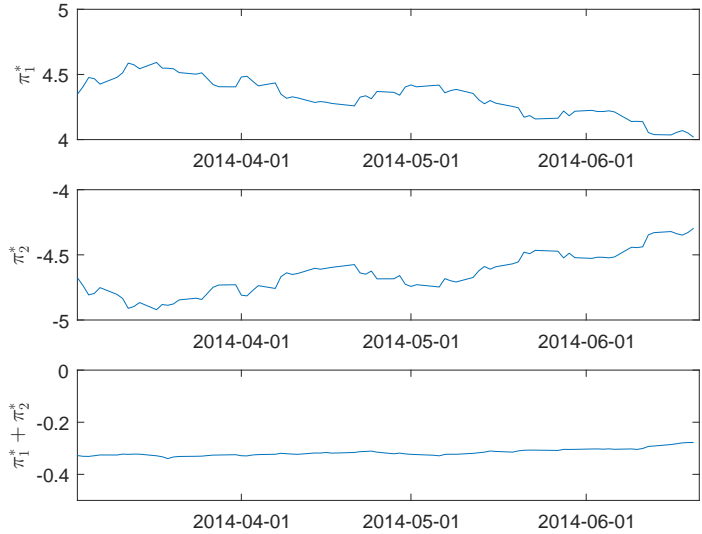}
\caption{
Optimal strategies  ${\pi}_1^*$, ${\pi}_2^*$ and ${\pi}_1^*+{\pi}_2^*$ based on historical  WTI crude oil futures data   over the period Mar 2014 - Jun 2014
using parameters as displayed in Table~\ref{s97table}.
}
\label{PLOT5}
\end{figure}

Having analyzed the parameter dependence of the optimal strategies in details, now we turn to their path behavior based on  historical data.
We consider the  June 2014 and  July 2014  WTI crude oil futures.  We show  the empirical  optimal positions over the period March 2014 to June 2014.
This period is chosen to correspond to the post-calibration period of \cite{Ewald2018}.  
Applying our the explicit formulae for the strategies, we compute ${\pi}_1^*$, ${\pi}_2^*$, and ${\pi}_1^*+{\pi}_2^*$ based on the daily  settlement prices of these contracts as well as the parameters in Table~\ref{s97table}. As shown in Figure~\ref{PLOT5}, the optimal strategy  ${\pi}_1^*$ is positive throughout this period, corresponding to a long position in the front-month contract, and the opposite holds for ${\pi}_2^*$.  
Taken together, the sum of both positions is negligibly small, corresponding to a net neutral position. 
Overall, the positions changed little when the parameters $\eta$ and $\bar{\eta}$ 
are kept fixed. The only variables that change are $F_i$ and $T_i-t$, 
of which we have already seen the relative insensitivity in Figure~\ref{PLOT2}.

\begin{figure}[h]
    \centering
\includegraphics[width=4.2in]{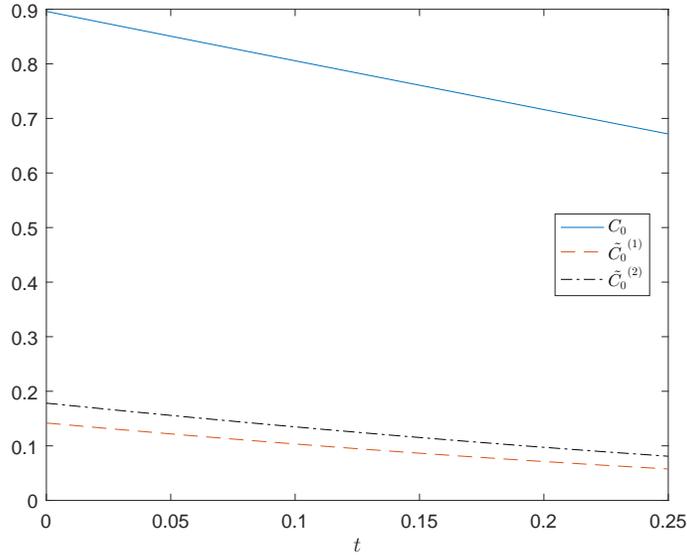}
\caption{
The certainty equivalents 
$C_0$ for the two-futures portfolio, as well as $\tilde{C}_0^{(1)}$ and $\tilde{C}_0^{(2)}$ for the single-futures portfolios, respectively with $T_1$-futures and $T_2$-futures (see \eqref{cetilde}). The certainty equivalents are evaluated at time $t=0$ with initial wealth  $w=0$. The trading  horizon is $T=1$, maturity of $F_1$ is $T_1=13/12$, and maturity of $T_2=14/12$. Other common parameters are from Table \ref{s97table}, along with  $F_1=100$ and $F_2=100$.
}
\label{PLOT6}
\end{figure}

We now turn  our attention to the  certainty equivalents.  With reference to Section \ref{sect_ce}, we    plot   in Figure \ref{PLOT6} the following certainty equivalents: 
$\tilde{C}^{(1)}$ in the single-futures case with $T_1$-futures traded, $\tilde{C}^{(2)}$ in the single-futures case with $T_2$-futures traded, and $C$ in the two-futures case with $T_1$-futures and $T_2$ futures traded. Their numerical values  are given in Table~\ref{cetable}.

\vspace{5pt}
\begin{table}[h]
\centering
{\setlength\extrarowheight{5pt}
$\begin{tabu}{|c|c|c|}
\hline
C_0(0) & \tilde{C}^{(1)}_0(0) & \tilde{C}^{(2)}_0(0) \\ \hline 
0.8962 &   0.1418 &   0.1782 \\ \hline
\end{tabu}$}\caption{Values of certainty equivalent: 
$\tilde{C}^{(1)}$ in the single-futures case with $T_1$-futures traded, $\tilde{C}^{(2)}$ in the single-futures case with $T_2$-futures traded, and $C$ in the two-futures case with $T_1$-futures and $T_2$ futures traded. The certainty equivalents are evaluated at $t=0$ and $w=0$.}
\label{cetable}
\end{table}
\vspace{5pt}
\newpage

%

We observe from Figure \ref{PLOT6}  that the certainty equivalent for trading  two contracts simultaneously is significantly greater than that  derived from trading only a single contract regardless of the choice of maturity.  
In fact, the certainty equivalent $C$ is much larger than the sum of the two certainty equivalents $\tilde{C}^{(1)}$ and $\tilde{C}^{(2)}$. This makes sense since   the single-contract case can be viewed as two-contracts case but with one strategy constrained at zero. Effectively, the single-contract case is  restricting the admissible set from  $\mathcal{A}_t$ to $\tilde{\mathcal{A}}_t$, thus reducing the maximum expected utility as well as   the certainty equivalent.
Our result confirms the intuition that more choices of trading instruments  are  preferable to fewer.

\begin{figure}[h]
    \centering
{\includegraphics[width=4.2in]{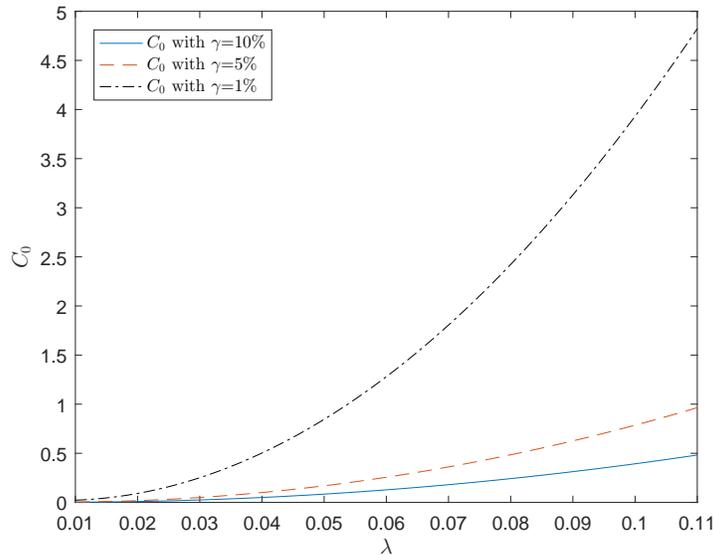}}
\caption{
Certainty equivalent $C_0$, at time $t=0$  with zero initial wealth $W_0=0$,
 as a function of the market price of risk $\lambda$,  
with parameters as displayed in Table~\ref{s97table}.
}
\label{PLOT7}
\end{figure}

Lastly, we examine the behavior of $C$ at different risk aversion levels with focus on its sensitivity with respect to  the market price of risk $\lambda$. In Figure \ref{PLOT7}, we  see that  the certainty equivalent at time 0,   $C_0$, is  increasing and  quadratic in $\lambda$, 
and tends to infinity as $\lambda$ increases. This holds for all three values of $\gamma$ shown, but a lower risk aversion suggests that the certainty equivalent  is higher and faster growing in $\lambda$.


\clearpage

\section{Conclusions}
\label{conclusions}
We have analyzed the problem of dynamically trading two futures contracts with the same underlying.  Under a  two-factor mean-reverting    model for the spot price,  we derive  the futures price dynamics  and solve  the portfolio optimization problem in closed form and give explicit optimal  trading strategies.   By studying  the associated Hamilton-Jacobi-Bellman equation, we solve the utility maximization explicitly and provide the optimal  trading strategies  in closed form. In addition to  the analytic properties of our solutions, we also apply our results to commodity futures trading and   present numerical examples to illustrate the optimal holdings.

There are several natural directions for future research on managed futures. First, additional factors and sources of risks can be incorporated in the spot model, including random jumps,  stochastic volatility, and stochastic interest rate. Nevertheless,   more complex models typically mean that the value function and optimal trading strategies are not available in closed form and  thus  require numerical approximations. In reality, futures are typically traded with leverage, and margin requirement is a core issue. Incorporating this feature to futures portfolio optimization may not be straightforward, but will certainly have practical implications.

\appendix
\section{Appendix}
\label{append}
\subsection{Portfolio with Three Futures Contracts}
Let us consider a dynamic  portfolio of three futures contracts with different maturities $T_1, T_2$ and $T_3$. In this case, the  wealth  and  futures prices follow the system of SDEs
\[
\begin{bmatrix} d{W}_t \\ d{F}^{(1)}_t \\ d{F}^{(2)}_t \\ d{F}^{(3)}_t \end{bmatrix}=
\begin{bmatrix} 
{\pi}_1 {\mu}_1(t) {F}^{(1)}_t + 
{\pi}_2 {\mu}_2(t) {F}^{(2)}_t+ 
{\pi}_3 {\mu}_3(t) {F}^{(3)}_t
\\ {\mu}_1(t) {F}^{(1)}_t \\ {\mu}_2(t) {F}^{(2)}_t \\ {\mu}_3(t) {F}^{(3)}_t \end{bmatrix}
dt
\]
\begin{equation}
+
\begin{bmatrix} 
{\pi}_1 \eta {F}^{(1)}_t+ 
{\pi}_2 \eta {F}^{(2)}_t + 
{\pi}_3 \eta {F}^{(3)}_t 
& {\pi}_1 \bar{\eta}B_1(t) {F}^{(1)}_t+ 
{\pi}_2 \bar{\eta}B_2(t) {F}^{(2)}_t+ 
{\pi}_3 \bar{\eta}B_3(t) {F}^{(3)}_t
\\
\eta {F}^{(1)}_t & \bar{\eta}B_1(t) {F}^{(1)}_t\\
\eta {F}^{(2)}_t & \bar{\eta}B_2(t) {F}^{(2)}_t\\
\eta {F}^{(3)}_t & \bar{\eta}B_3(t) {F}^{(3)}_t\\
\end{bmatrix}
\begin{bmatrix} dZ^s_t \\ dZ^\delta_t \end{bmatrix}.
\label{system3}
\end{equation}


The    HJB equation associated with  the value function $u(t,w,F_1,F_2,F_3)$ is 
\begin{align}\nonumber 
u_t &+ \sup_{\pi_1,\pi_2,\pi_3} \left[\begin{array}{c} \end{array} \right.
\pi_1 \mu_1 F_1 u_w + \pi_2 \mu_2  F_2 u_w  + \pi_3 \mu_3  F_3 u_w \\\nonumber 
&+ F_1 (\pi_1 \sigma_1^2 F_1 
+\pi_2 (\eta^2+\bar{\eta}^2 B_1B_2+\rho\eta\bar{\eta}(B_1+B_2)) F_2
+\pi_3 (\eta^2+\bar{\eta}^2 B_1B_3+\rho\eta\bar{\eta}(B_1+B_3)) F_3) u_{w1} \\\nonumber
&+ F_2 (
 \pi_1 (\eta^2+\bar{\eta}^2B_1B_2+\rho\eta\bar{\eta}(B_1+B_2)) F_1
+\pi_2 {\sigma}_2^2 F_2 
+\pi_3 (\eta^2+\bar{\eta}^2B_2B_3+\rho\eta\bar{\eta}(B_2+B_3)) F_3) u_{w2} \\\nonumber
&+ F_3 (
\pi_1 (\eta^2+\bar{\eta}^2B_1 B_3+\rho\eta\bar{\eta}(B_1+B_3)) F_1+
\pi_2 (\eta^2+\bar{\eta}^2B_2 B_3+\rho\eta\bar{\eta}(B_2+B_3)) F_2
+\pi_3 {\sigma}_3^2 F_3) u_{w3} \\\nonumber
&+\frac{1}{2}(F_1^2 \pi_1^2 \sigma_1^2 + F_2^2 \pi_2^2 \sigma_2^2 + F_3^2 \pi_3^2 \sigma_3^2 
 +2(\eta^2+\bar{\eta}^2 B_1 B_2+\rho\eta\bar{\eta}(B_1+B_2))F_1 F_2 \pi_1\pi_2 \\\nonumber
&+2(\eta^2+\bar{\eta}^2 B_1 B_3+\rho\eta\bar{\eta}(B_1+B_3))F_1 F_3 \pi_1\pi_3\\\nonumber
&+2(\eta^2+\bar{\eta}^2 B_2 B_3+\rho\eta\bar{\eta}(B_2+B_3))F_2 F_3 \pi_2\pi_3
)u_{ww} \left.  \begin{array}{c} \end{array}\right] \\ \nonumber
&+\frac{\sigma_1^2}{2} F_1^2 u_{11} + \frac{\sigma_2^2}{2} F_2^2 u_{22} + \frac{\sigma_3^2}{2} F_3^2 u_{33} + \mu_1 F_1 u_1 +\mu_2 F_2 u_2 +\mu_3 F_3 u_3 \\ \nonumber
&+ (\eta^2+\bar{\eta}^2 B_1 B_2+\rho\eta\bar{\eta}(B_1+B_2)) F_1 F_2 {u}_{12} 
 + (\eta^2+\bar{\eta}^2 B_1 B_3+\rho\eta\bar{\eta}(B_1+B_3)) F_1 F_3 {u}_{13} \\ \nonumber
&+ (\eta^2+\bar{\eta}^2 B_2 B_3+\rho\eta\bar{\eta}(B_2+B_3)) F_2 F_3 {u}_{23} = 0,
\end{align}
where we suppress the dependence on $t$  in $\mu_i(t), \sigma_i(t)$ and $B_i(t)$, for $i=1,2,3$.

To solve for the optimal strategies ($\pi_1,\pi_2,\pi_3$), we  impose  the first-order conditions. 
To facilitate the presentation, we define the constants
\[
a_{ij} \equiv \eta^2+\bar{\eta}^2B_iB_j+\rho\eta\bar{\eta}(B_i+B_j), \quad i,j = 1,2,3.
\]
This leads to the following system of equations
\begin{equation}u_{ww}
\begin{bmatrix} 
F_1^2  \sigma_1^2 & F_1 F_2 a_{12} & F_1 F_3 a_{13} \\
F_1 F_2 a_{13} & F_2^2  \sigma_2^2 & F_2 F_3  a_{23} \\
F_1 F_3  a_{13} & F_2 F_3  a_{23} & F_3^2  \sigma_3^2\\
\end{bmatrix}
\begin{bmatrix} \pi_1 \\ \pi_2 \\ \pi_3 \end{bmatrix}
=-
\begin{bmatrix}
F_1 u_w \mu_1 +F_1^2 u_{w1}\sigma_1^2 +F_1 F_2 u_{w2}a_{12} +F_1 F_3 u_{w3}a_{13} \\ 
F_2 u_w \mu_2 +F_1 F_2 u_{w1}a_{12} +F_2^2 u_{w2}\sigma_2^2 +F_2 F_3 u_{w3}a_{23} \\ 
F_3 u_w \mu_3 +F_1 F_3 u_{w1}a_{13} +F_2 F_3 u_{w2}a_{23} +F_3^2 u_{w3}\sigma_3^2 \\ 
\end{bmatrix},
\end{equation}
which is singular as verified by computation.

\clearpage\begin{small}
\begin{spacing}{0.7}
\bibliographystyle{apa}
\bibliography{biblio}
\end{spacing}
\end{small}
\end{document}